\documentclass[         %
aps,                    
jpa,                    
showpacs,               
nofootinbib,            
showkeys,               %
preprintnumbers,        %
floatfix]               
{revtex4}               
\tighten
\usepackage{graphicx} 

\begin{document}

\title{Neutrinos in Astrophysics and Cosmology}


\author{A.B. Balantekin} 
\address{Physics Department, University of Wisconsin, Madison WI 53706 USA}

\begin{abstract}
Neutrinos play a crucial role in many aspects of astrophysics and cosmology. Since they control the electron fraction, or equivalently neutron-to-proton ratio, neutrino properties impact yields of r-process nucleosynthesis. Similarly the weak decoupling temperature in the Big Bang Nucleosynthesis epoch is exponentially dependent on the neutron-to-proton ratio. In these conference proceedings, I briefly summarize some of the recent work exploring the role of neutrinos in astrophysics and cosmology. 
\end{abstract}

\maketitle


\section{INTRODUCTION}

Revolutionary advances are taking place in astrophysics and cosmology motivated by the precision instruments such as COBE, WMAP, Planck, LIGO and powerful telescopes such as Subaru and Keck. State-of-the-art nuclear and particle physics facilities complement these powerful tools for the study of our Universe. An assessment of status 
of physical  sciences at the beginning of the new milenium is given by the U.S. National Academy of Sciences in the report entitled ``Connecting Quarks with the Cosmos: Eleven  Science Questions For the New Century". One of the outstanding questions was about neutrinos: What are the masses of neutrinos, and how have they shaped the evolution of the Universe?  In this talk I will discuss aspects of this quest. 

A related question, which does not seem to be related to neutrino physics at the first sight, is the origin of all the elements around us. We know that the Big Bang Nucleosynthesis (BBN) produces hydrogen, helium, and a very small amount of light elements such deuterium, beryllium, and lithium. Lack of stable nuclei with $A=5$ and $A=8$ as well as rapidly rising Coulomb barriers after those bottlenecks prevents BBN from producing any other nuclei. These light elements condense into population III stars, which are very big and obviously very poor in metals (astronomers's lingo for anything that is not hydrogen and helium). These stars rapidly burn their nuclear fuel, produce many of the elements up to and including most-stable nuclei near iron. When they go supernova these stars then populate the Universe with a scattering of iron group nuclei. From their ashes, population II stars, which are still rather metal poor, are formed. Some of them go supernova and produce elements such as U. Eu, Th,.. via the rapid neutron capture (r-process). The so-called AGB stars produce elements such as Ba, La,, Y, ... via the slow neutron capture (s-process). Here the terms "rapid" or "slow" are in comparison to the weak interaction timescale.

This is a neat picture, but it has many missing pieces. Abundances of the $A=7$ nuclei, calculated with the state-of-the-art nuclear physics techniques, are 30\% off the observed values.  Despite many attempts, a satisfactory explanation of this deficit is still lacking \cite{Cyburt:2015mya,Coc:2014oia,Iocco:2008va,Steigman:2007xt}. 
Similarly we do not know the astrophysical site of the r-process nucleosynthesis \cite{Arcones:2016euo,Qian:2007vq}. 
Several sites have been considered \cite{Hansen:2014tfa}. 
Among the leading candidates are neutron-star mergers and core-collapse supernovae. As I illustrate in the next paragraph neutrinos not only play a crucial role in the dynamics of these sites, but they also control the value of the electron fraction, the parameter determining yields of the nucleosynthesis processes \cite{Balantekin:2003ip}. 

For a neutral medium with only electrons, protons and neutrons, the electron fraction is given by  
\begin{equation}
\label{1.1}
Y_e = \frac{N_p}{N_n+N_p}
\end{equation}
and satisfies the differential equation
\begin{equation}
\label{1.2}
\frac{dY_e}{dt} = \lambda_n -\left(\lambda_p+\lambda_n \right) Y_e,
\end{equation}
where $\lambda_p$ is the proton weak loss rate, i.e. the rate for the 
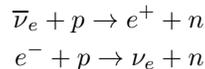
\begin{eqnarray*}
\overline{\nu}_e + p \rightarrow e^+ + n \\
e^-+p \rightarrow \nu_e +n
\end{eqnarray*}
reactions. Similarly $\lambda_n$ is the neutron weak loss rate, i.e. the rate for the 
\begin{eqnarray*}
\nu_e + n \rightarrow e^- + p \\
e^++n \rightarrow \overline{\nu}_e +p
\end{eqnarray*}
reactions. Equilibrium value for the $Y_e$ is obtained by setting its derivative to zero. One gets 
\begin{equation}
\label{1.2a}
Y_e^{(0)} = \frac{\lambda_n}{\lambda_p+\lambda_n}
\end{equation}
where the superscript zero indicates that this equilibrium value is only for a mixture which does not include any other hadrons besides proton and neutron. A robust r-process nucleosynthesis requires rather small values of the electron fraction. The rate for the neutrino induced reactions is 
\begin{equation}
\label{1.3}
\lambda = \int \sigma (E_{\nu}) \frac{d \phi_{\nu}}{dE_{\nu}} dE_{\nu}.
\end{equation}
Clearly not only the equilibrium value of the electron fraction but all pre-equilibrium values depend very sensitively on the relative amounts of the electron neutrino and electron antineutrino fluxes. These amounts are in turn controlled by the neutrino properties. In the rest of this talk I outline how particular neutrino physics components impact various astrophysical sites. For further discussion the reader is referred to Ref. \cite{Balantekin:2013gqa}. 

\section{Collective neutrino oscillations}

Recent development of two- and and three-dimensional models for core-collapse supernovae unveil a complex interplay between turbulence, neutrino physics and thermonuclear reactions. Understanding a core-collapse supernova requires answers to a variety of questions some of which need to be answered by neutrino physics, both theoretically and experimentally. Knowledge of high-density equation of state and neutrino interactions are crucial to investigate the proto-neutron star. Propagation of neutrinos after they decouple from the neutrinosphere determines to what extent they contribute to shock reheating and various nucleosynthesis scenarios. Finally detection of the supernova neutrino burst requires a sufficiently precise knowledge of neutrino interactions with target nuclei, such as $^{40}$Ar, in the current and planned detectors.

The progenitor star is an order of magnitude heavier than the Sun. Its core-collapse releases about $10^{59}$ MeV of energy, $99 \%$ of which is carried out by neutrinos of all flavors. Since each neutrino carries out on the average $1 \sim 10$ MeV of energy one expects a total number of $10^{57} \sim 10^{58}$ neutrinos to be released. This necessitates including the effects of neutrino-neutrino interactions in neutrino transport 
\cite{Duan:2009cd,Duan:2010bg,Raffelt:2010zza}. The resulting phenomena is termed "collective neutrino oscillations"  and could impact supernova nucleosynthesis \cite{Balantekin:2004ug,Duan:2010af}. It should be pointed out that the phenomenology of such oscillations is likely to be very rich.  For example, merger of neutron stars with other neutron stars or black holes produce, unlike core-collapse supernovae, more antineutrinos than neutrinos. In such cases matter-neutrino resonances are possible \cite{Malkus:2014iqa}.

To emphasize many-body aspects of the neutrino gas in a core-collapse supernova we 
introduce the creation and annihilation operators for a neutrino 
with three momentum ${\bf p}$, and write down the generators of the neutrino flavor isospin algebras with two flavors   
\cite{Balantekin:2006tg}: 
\begin{eqnarray}
J_+({\bf p}) &=& a_x^\dagger({\bf p}) a_e({\bf p}), \> \> \>
J_-({\bf p})=a_e^\dagger({\bf p}) a_x({\bf p}), \nonumber \\
J_0({\bf p}) &=& \frac{1}{2}\left(a_x^\dagger({\bf p})a_x({\bf p})-a_e^\dagger({\bf p})a_e({\bf p})
\right). \label{su2}
\end{eqnarray}
Using the operators in Eq. (\ref{su2}) 
the Hamiltonian for a neutrino propagating through matter and interacting with other neutrinos takes the form  
\cite{Balantekin:2006tg} 
\begin{equation}
\label{total}
H = H_{\nu} + H_{\nu \nu} 
= \left(
\sum_p\frac{\delta m^2}{2p}\hat{B}\cdot\vec{J}_p  - \sqrt{2} G_F 
N_e  J_p^0  \right) 
+ \frac{\sqrt{2}G_{F}}{V}\sum_{\mathbf{p},\mathbf{q}}\left(1- 
\cos\vartheta_{\mathbf{p}\mathbf{q}}\right)\vec{J}_{\mathbf{p}}\cdot\vec{J}_{\mathbf{q}}  
\end{equation} 
where the auxiliary vector quantity $\hat{B}$ is given by 
\begin{equation}
\hat{B} = (\sin2\theta,0,-\cos2\theta) ,
\end{equation}
$N_e$ is the background electron density and $\delta m^2$ is the difference between squares of the masses associated with mass eigenstates. As we see below, introducing this fictitious "magnetic field" helps analogies with the condensed-matter problems. In the above equations $\theta$ is the mixing angle between electron 
neutrino and the other neutrino flavor and $\vartheta_{\bf pq}$  is the angle between neutrino momenta ${\bf p}$ and {\bf q}.  This Hamiltonian assumes an adiabatic expansion of the many-neutrino gas and gives rise to neutrino collective oscillations.  For the ease of presentation in the above discussion we only consider neutrinos of two flavors and omit antineutrinos. However, one can easily include antineutrinos (by introducing a second set of SU(2) algebras) and three flavors (by using SU(3) algebras for both neutrinos and antineutrinos) in this formalism. 

In the limit one can average over the angles $\vartheta_{\bf pq}$ between the momenta of the interacting neutrinos ("the single angle limit") the Hamiltonian in Eq. (\ref{total}) takes the form
\begin{equation}
H = \sum_p \frac{\delta m^2}{2p}\hat{B}\cdot\vec{J}_p  + 2 \mu \vec{J}_\cdot \vec{J}
\end{equation}
where $\vec{J} =\sum_{\mathbf{p}} \vec{J}_{\mathbf{p}}$ and $\mu =  \frac{G_F}{\sqrt{2}V} <1-cos \vartheta>$. Aside from the sign of the second term this is the same as the BCS Hamiltonian of superconductivity. Hence in the single-angle limit, the dynamics of the neutrino collective oscillations is analogous to that of the BCS Hamiltonian. Just like the BCS case, eigenstates and eigenvalues of the collective neutrino oscillation Hamiltonian can be found using the Bethe ansatz technique \cite{Pehlivan:2011hp}. There are also invariants of the motion:
\begin{equation}
h_p = J^0_p + 2 \mu \sum_{p,q, p\neq q} \frac{\vec{J}_p\cdot \vec{J}_q}{\delta m^2 \left(\frac{1}{p}-\frac{1}{q} \right)} .
\end{equation}
These dynamical symmetries are present even when the CP symmetry is violated in neutrino oscillations 
\cite{Pehlivan:2014zua}. Aside from those symmetries neutrino-neutrino interactions lead to novel collective and emergent effects, such as conserved quantities and interesting features in the neutrino energy spectra called spectral "swaps" or "splits" \cite{Raffelt:2007cb,Duan:2008za,Galais:2011gh}. An excat solution of the problem seems to be currently out of reach, however adiabatic solutions of the exact many-body Hamiltonian are explored in some limited cases \cite{Pehlivan:2016lxx}. 

Beyond the single-angle approximation a commonly used approach is to introduce a mean field reducing the two-body problem into a one-body problem: 
\begin{equation}
a^{\dagger}a^{\dagger} aa \Rightarrow \langle a^{\dagger} a \rangle a^{\dagger}a + \langle a^{\dagger} a^{\dagger} \rangle aa + {\mathrm H.c.}
\end{equation}
where the averages are calculated using the mean-field state. In this way, for example for Majorana neutrinos, one can reduce the neutrino-neutrino interaction 
\begin{equation}
\overline{\psi}_{\nu L} \gamma^{\mu} \psi_{\nu L}\overline{\psi}_{\nu L} \gamma_{\mu} \psi_{\nu L} \Rightarrow 
\overline{\psi}_{\nu L} \gamma^{\mu} \psi_{\nu L} \langle \overline{\psi}_{\nu L} \gamma_{\mu} \psi_{\nu L} \rangle + \cdots ,
\end{equation}
the antineutrino-antineutrino interaction 
\begin{equation}
\overline{\psi}_{\nu R} \gamma^{\mu} \psi_{\nu R}\overline{\psi}_{\nu R} \gamma_{\mu} \psi_{\nu R} \Rightarrow 
\overline{\psi}_{\nu R} \gamma^{\mu} \psi_{\nu R} \langle \overline{\psi}_{\nu R} \gamma_{\mu} \psi_{\nu R} \rangle + \cdots ,
\end{equation}
and the neutrino-antineutrino interaction
\begin{equation}
\overline{\psi}_{\nu L} \gamma^{\mu} \psi_{\nu L}\overline{\psi}_{\nu R} \gamma_{\mu} \psi_{\nu R} \Rightarrow 
\overline{\psi}_{\nu L} \gamma^{\mu} \psi_{\nu L} \langle \overline{\psi}_{\nu R} \gamma_{\mu} \psi_{\nu R} \rangle + \cdots .
\end{equation}
However, in addition to those one can also have a mean field comprised of both neutrino and antineutrino fields
\begin{equation}
\overline{\psi}_{\nu L} \gamma^{\mu} \psi_{\nu L}\overline{\psi}_{\nu R} \gamma_{\mu} \psi_{\nu R} \Rightarrow 
\overline{\psi}_{\nu L} \gamma^{\mu} \langle \psi_{\nu L}  \overline{\psi}_{\nu R} \gamma_{\mu} \rangle \psi_{\nu R}  + \cdots .
\end{equation}
Note that symmetry principles dictate such a mean field to be proportional to the neutrino mass. This mean field is negligible is the medium is isotropic, but can be important in a number of astrophysical sites where matter distribution is not isotropic 
\cite{Vlasenko:2013fja,Serreau:2014cfa,Cirigliano:2014aoa}. Once a mean field is introduced the resulting one-body Hamiltonian is amenable to the standard techniques. 

\section{Alpha effect and sterile neutrinos}

The expression for the electron fraction in the presence of alpha particles in addition to protons and neutrons in a neutral medium takes the form
\begin{equation}
\label{2.1}
Y_e = \frac{N_p+2N_{\alpha}}{N_p+N_n+N_{\alpha}}.
\end{equation}
Defining the mass fraction of the alpha particles 
\begin{equation}
\label{2.2}
X_{\alpha}= \frac{4N_{\alpha}}{N_p+N_n+4N_{\alpha}}
\end{equation}
one can write the evolution equation 
\begin{equation}
\label{2.3}
\frac{d}{dt} \left( Y_e - \frac{1}{2} X_{\alpha} \right) = \lambda_n -\left(\lambda_p+\lambda_n \right) Y_e +\frac{1}{2} 
\left(\lambda_p - \lambda_n \right) X_{\alpha}.
\end{equation}
Once the chemical equilibrium is reached the electron fraction takes the value
\begin{equation}
\label{ye}
Y_e = \frac{\lambda_n}{\lambda_n + \lambda_p} + \frac{1}{2} \left( \frac{\lambda_p - \lambda_n}{\lambda_p + \lambda_n} \right) X_{\alpha}. 
\end{equation}
One can rewrite Eq. (\ref{ye}) using Eq. (\ref{1.2a}) as 
\begin{equation}
Y_e =Y_e^{(0)} + \left( \frac{1}{2} - Y_e^{(0)} \right) X_{\alpha}. 
\end{equation}
Hence if $Y_e^{(0)}$ was less than one-half before the alpha particles are formed, electron fraction will increase after the introduction of the alpha particles. Similarly if $Y_e^{(0)}$ was more than one-half before, electron fraction will decrease after the introduction of the alpha particles. Hence the formation of alpha particles will force the electron fraction to get closer to one-half, inhibiting r-process nucleosynthesis. Electron neutrinos captured on neutrons convert them into protons. If the electron neutrino flux is high enough a significant number of protons thus created will catch two more neutrons and get locked into alpha particles. Since the alpha particle binding energy is very high, a large electron neutrino flux  could then prevent r-process nucleosynthesis from proceeding. This phenomenon is known as the alpha effect \cite{Meyer:1992zz,McLaughlin:1997qi}. 

One way to eliminate the alpha effect is to reduce the electron neutrino flux by allowing mixing of electron neutrinos with sterile neutrinos since sterile neutrinos do not take part in weak interactions \cite{McLaughlin:1999pd}. 
Indeed active-sterile mixing could yield very low values of the electron fraction \cite{McLaughlin:1999pd,Caldwell:1999zk,Fetter:2002xx}. Furthermore, the values of the sterile neutrino parameters which yield lowest value of the electron fraction is consistent with the neutrino parameters needed to explain the LSND results in terms of active-sterile neutrino mixing \cite{Abazajian:2012ys}. However, recent work indicates that active-sterile mixing with the parameters inferred from reactor anomaly \cite{Abazajian:2012ys} enables nucleosynthesis to proceed, but seems to suppress shock reheating by neutrinos \cite{Wu:2013gxa}. 

\section{Neutrino magnetic Moment effects in Big Bang Nucleosynthesis}

With input from nuclear physics and cosmology (especially the precision measurements of the baryon density) one can predict abundances of light elements formed in the Big Bang nucleosynthesis (BBN). Although d and $^4$He observations agree with the predictions, $^7$Li observations  are well below predictions. Considering nonthermal photons produced in the decay of the heavy sterile mass eigenstates due to the neutrino magnetic moment, one can explore the constraints imposed by the observed abundances of all the light elements produced during BBN. It is possible to provide an upper limit to the magnetic moments of massive sterile Dirac neutrinos which could fix the Li problem by imposing restrictions on the number of the additional helicity states and using the primordial helium abundance as a constraint \cite{Kusakabe:2013sna,Pospelov:2010hj}.

The production of light elements in BBN is exquisitely sensitive to the weak decoupling temperature since the neutron-to-proton ratio is exponentially dependent on it. Even if one ignores the possibility of the presence of sterile neutrinos,  
the extra couplings due to the magnetic interaction in addition to the usual weak interaction couplings alter the way active neutrinos decouple from the relativistic plasma of electrons/positrons and photons. However, the changes in cosmological parameters from magnetic contributions to neutrino decoupling temperatures are below the level of upcoming precision observations \cite{Vassh:2015yza}.

\section{Conclusions}

In this summary of a sampling of recent work exploring the role of neutrinos in astrophysics and cosmology a number of research directions are omitted. For example, one of the puzzles of the modern physics is the apparent matter-antimatter asymmetry in the Universe. Consequently stars and supernovae are not CP-invariant. Conditions under which additional CP-violation effects can be seen in supernovae are explored 
\cite{Pehlivan:2014zua,Balantekin:2007es,Gava:2008rp}. Another example is neutrino propagation in noisy, turbulent, or stochastic media \cite{Loreti:1994ry}. Such situations surely exist in astrophysical sites. Various authors investigated effects of noisy background matter, stochasticity and turbulence in the Sun \cite{Balantekin:2003qm} and in core-collapse supernovae 
\cite{Loreti:1995ae,Patton:2014lza}.

\section*{Acknowledgments}
This work was supported in part 
by the U.S. National Science Foundation Grant No.  PHY-1514695 
and
in part by the University of Wisconsin Research Committee with funds
granted by the Wisconsin Alumni Research Foundation. 
I also thank the Center for Theoretical Underground Physics and Related Areas (CETUP* 2015) in South Dakota for its hospitality and for partial support during the completion of this work. 



\begin{thebibliography}{99}

\bibitem{Cyburt:2015mya} 
  R.~H.~Cyburt, B.~D.~Fields, K.~A.~Olive and T.~H.~Yeh,
  arXiv:1505.01076 [astro-ph.CO].

\bibitem{Coc:2014oia} 
  A.~Coc, J.~P.~Uzan and E.~Vangioni,
  JCAP {\bf 1410}, 050 (2014)
  [arXiv:1403.6694 [astro-ph.CO]].
  
\bibitem{Iocco:2008va} 
  F.~Iocco, G.~Mangano, G.~Miele, O.~Pisanti and P.~D.~Serpico,
  Phys.\ Rept.\  {\bf 472}, 1 (2009)
  [arXiv:0809.0631 [astro-ph]].
  
\bibitem{Steigman:2007xt} 
  G.~Steigman,
  Ann.\ Rev.\ Nucl.\ Part.\ Sci.\  {\bf 57}, 463 (2007)
  [arXiv:0712.1100 [astro-ph]].

\bibitem{Arcones:2016euo} 
  A.~Arcones {\it et al.},
  arXiv:1603.02213 [astro-ph.SR].

\bibitem{Qian:2007vq} 
  Y.-Z.~Qian and G.~J.~Wasserburg,
  Phys.\ Rept.\  {\bf 442}, 237 (2007)
  doi:10.1016/j.physrep.2007.02.006
  [arXiv:0708.1767 [astro-ph]].

\bibitem{Hansen:2014tfa} 
  C.~J.~Hansen, F.~Montes and A.~Arcones,
  Astrophys.\ J.\  {\bf 797}, no. 2, 123 (2014)
  doi:10.1088/0004-637X/797/2/123
  [arXiv:1408.4135].

\bibitem{Balantekin:2003ip} 
   See e.g. 
  A.~B.~Balantekin and G.~M.~Fuller,
  J.\ Phys.\ G {\bf 29}, 2513 (2003)
  [astro-ph/0309519].

\bibitem{Balantekin:2013gqa} 
  A.~B.~Balantekin and G.~M.~Fuller,
  Prog.\ Part.\ Nucl.\ Phys.\  {\bf 71}, 162 (2013)
  doi:10.1016/j.ppnp.2013.03.008
  [arXiv:1303.3874 [nucl-th]].

\bibitem{Duan:2009cd} 
  H.~Duan and J.~PKneller,
  \emph{J.\ Phys.\ G} {\bf 36}, 113201 (2009)
  [arXiv:0904.0974 [astro-ph.HE]].

\bibitem{Duan:2010bg} 
  H.~Duan, G.~M.~Fuller and Y.~-Z.~Qian,
  \emph{Ann.\ Rev.\ Nucl.\ Part.\ Sci.}  {\bf 60}, 569 (2010)
  [arXiv:1001.2799 [hep-ph]].

\bibitem{Raffelt:2010zza} 
  G.~G.~Raffelt,
  \emph{Prog.\ Part.\ Nucl.\ Phys.}  {\bf 64}, 393 (2010).

\bibitem{Balantekin:2004ug} 
  A.~B.~Balantekin and H.~Yuksel,
  \emph{New J.\ Phys.}  {\bf 7}, 51 (2005)
  [astro-ph/0411159].

\bibitem{Duan:2010af} 
  H.~Duan, A.~Friedland, G.~C.~McLaughlin and R.~Surman,
  \emph{J.\ Phys.\ G} {\bf 38}, 035201 (2011)
  [arXiv:1012.0532 [astro-ph.SR]].

\bibitem{Malkus:2014iqa} 
  A.~Malkus, A.~Friedland and G.~C.~McLaughlin,
  arXiv:1403.5797 [hep-ph].

\bibitem{Balantekin:2006tg} 
  A.~B.~Balantekin and Y.~Pehlivan,
  J.\ Phys.\ G {\bf 34}, 47 (2007)
  doi:10.1088/0954-3899/34/1/004
  [astro-ph/0607527].
  
\bibitem{Pehlivan:2011hp} 
  Y.~Pehlivan, A.~B.~Balantekin, T.~Kajino and T.~Yoshida,
  Phys.\ Rev.\ D {\bf 84}, 065008 (2011)
  doi:10.1103/PhysRevD.84.065008
  [arXiv:1105.1182 [astro-ph.CO]].
  
\bibitem{Pehlivan:2014zua} 
  Y.~Pehlivan, A.~B.~Balantekin and T.~Kajino,
  Phys.\ Rev.\ D {\bf 90}, no. 6, 065011 (2014)
  doi:10.1103/PhysRevD.90.065011
  [arXiv:1406.5489 [hep-ph]].
  
\bibitem{Raffelt:2007cb} 
  G.~G.~Raffelt and A.~Y.~Smirnov,
  \emph{Phys.\ Rev.\ D} {\bf 76}, 081301 (2007)
  [Erratum-ibid.\ D {\bf 77}, 029903 (2008)]
  [arXiv:0705.1830 [hep-ph]].

\bibitem{Duan:2008za} 
  H.~Duan, G.~M.~Fuller and Y.~-Z.~Qian,
  \emph{Phys.\ Rev.\ D} {\bf 77}, 085016 (2008)
  [arXiv:0801.1363 [hep-ph]].

\bibitem{Galais:2011gh} 
  S.~Galais and C.~Volpe,
  Phys.\ Rev.\ D {\bf 84}, 085005 (2011)
  [arXiv:1103.5302 [astro-ph.SR]].

\bibitem{Pehlivan:2016lxx} 
  Y.~Pehlivan, A.~L.~Subasi, N.~Ghazanfari, S.~Birol and H.~Yüksel,
  arXiv:1603.06360 [astro-ph.HE].

\bibitem{Vlasenko:2013fja} 
  A.~Vlasenko, G.~M.~Fuller and V.~Cirigliano,
  Phys.\ Rev.\ D {\bf 89}, no. 10, 105004 (2014)
  doi:10.1103/PhysRevD.89.105004
  [arXiv:1309.2628 [hep-ph]].

\bibitem{Serreau:2014cfa} 
  J.~Serreau and C.~Volpe,
  Phys.\ Rev.\ D {\bf 90}, no. 12, 125040 (2014)
  doi:10.1103/PhysRevD.90.125040
  [arXiv:1409.3591 [hep-ph]].

\bibitem{Cirigliano:2014aoa} 
  V.~Cirigliano, G.~M.~Fuller and A.~Vlasenko,
  Phys.\ Lett.\ B {\bf 747}, 27 (2015)
  doi:10.1016/j.physletb.2015.04.066
  [arXiv:1406.5558 [hep-ph]].

\bibitem{Meyer:1992zz} 
  B.~S.~Meyer, G.~J.~Mathews, W.~M.~Howard, S.~E.~Woosley and R.~D.~Hoffman,
  Astrophys.\ J.\  {\bf 399}, 656 (1992).
  doi:10.1086/171957

\bibitem{McLaughlin:1997qi} 
  G.~C.~McLaughlin, G.~M.~Fuller and J.~R.~Wilson,
  Astrophys.\ J.\  {\bf 472}, 440 (1996)
  doi:10.1086/178077
  [astro-ph/9701114].

\bibitem{McLaughlin:1999pd} 
  G.~C.~McLaughlin, J.~M.~Fetter, A.~B.~Balantekin and G.~M.~Fuller,
  Phys.\ Rev.\ C {\bf 59}, 2873 (1999)
  doi:10.1103/PhysRevC.59.2873
  [astro-ph/9902106].

\bibitem{Caldwell:1999zk} 
  D.~O.~Caldwell, G.~M.~Fuller and Y.~Z.~Qian,
  Phys.\ Rev.\ D {\bf 61}, 123005 (2000)
  doi:10.1103/PhysRevD.61.123005
  [astro-ph/9910175].

\bibitem{Fetter:2002xx} 
  J.~Fetter, G.~C.~McLaughlin, A.~B.~Balantekin and G.~M.~Fuller,
  Astropart.\ Phys.\  {\bf 18}, 433 (2003)
  doi:10.1016/S0927-6505(02)00156-1
  [hep-ph/0205029].

\bibitem{Abazajian:2012ys} 
  K.~N.~Abazajian {\it et al.},
  arXiv:1204.5379 [hep-ph].

\bibitem{Wu:2013gxa} 
  M.~R.~Wu, T.~Fischer, L.~Huther, G.~Martínez-Pinedo and Y.~Z.~Qian,
  Phys.\ Rev.\ D {\bf 89}, no. 6, 061303 (2014)
  doi:10.1103/PhysRevD.89.061303
  [arXiv:1305.2382 [astro-ph.HE]].

\bibitem{Kusakabe:2013sna} 
  M.~Kusakabe, A.~B.~Balantekin, T.~Kajino and Y.~Pehlivan,
  Phys.\ Rev.\ D {\bf 87}, no. 8, 085045 (2013)
  doi:10.1103/PhysRevD.87.085045
  [arXiv:1303.2291 [astro-ph.CO]].
  
\bibitem{Pospelov:2010hj} 
  M.~Pospelov and J.~Pradler,
  Ann.\ Rev.\ Nucl.\ Part.\ Sci.\  {\bf 60}, 539 (2010)
  doi:10.1146/annurev.nucl.012809.104521
  [arXiv:1011.1054 [hep-ph]].

\bibitem{Vassh:2015yza} 
  N.~Vassh, E.~Grohs, A.~B.~Balantekin and G.~M.~Fuller,
  Phys.\ Rev.\ D {\bf 92}, no. 12, 125020 (2015)
  doi:10.1103/PhysRevD.92.125020
  [arXiv:1510.00428 [astro-ph.CO]].

\bibitem{Balantekin:2007es} 
  A.~B.~Balantekin, J.~Gava and C.~Volpe,
  Phys.\ Lett.\ B {\bf 662}, 396 (2008)
  doi:10.1016/j.physletb.2008.03.038
  [arXiv:0710.3112 [astro-ph]].

\bibitem{Gava:2008rp} 
  J.~Gava and C.~Volpe,
  Phys.\ Rev.\ D {\bf 78}, 083007 (2008)
  doi:10.1103/PhysRevD.78.083007
  [arXiv:0807.3418 [astro-ph]].

\bibitem{Loreti:1994ry} 
  F.~N.~Loreti and A.~B.~Balantekin,
  Phys.\ Rev.\ D {\bf 50}, 4762 (1994)
  doi:10.1103/PhysRevD.50.4762
  [nucl-th/9406003].

\bibitem{Balantekin:2003qm} 
  A.~B.~Balantekin and H.~Yuksel,
  Phys.\ Rev.\ D {\bf 68}, 013006 (2003)
  doi:10.1103/PhysRevD.68.013006
  [hep-ph/0303169].

\bibitem{Loreti:1995ae} 
  F.~N.~Loreti, Y.~Z.~Qian, G.~M.~Fuller and A.~B.~Balantekin,
  Phys.\ Rev.\ D {\bf 52}, 6664 (1995)
  doi:10.1103/PhysRevD.52.6664
  [astro-ph/9508106].

\bibitem{Patton:2014lza} 
  K.~M.~Patton, J.~P.~Kneller and G.~C.~McLaughlin,
  Phys.\ Rev.\ D {\bf 91}, no. 2, 025001 (2015)
  doi:10.1103/PhysRevD.91.025001
  [arXiv:1407.7835 [hep-ph]].

\end{thebibliography}

\end{document}